\documentclass[]{article}
\usepackage{graphicx}
\usepackage{amsmath, amsfonts, amsthm, amssymb, amstext, wrapfig, caption, , float, multirow, tabularx,bm,ragged2e,array,scrextend, verbatim,enumerate,mathtools,indentfirst}
\usepackage[colorlinks]{hyperref}
\hypersetup{
	citecolor = blue
}
\usepackage{authblk}
\usepackage{xcolor}

\theoremstyle{definition}
\usepackage[linesnumbered,ruled,vlined]{algorithm2e}
\SetKwInput{KwInput}{Input}                
\SetKwInput{KwOutput}{Output}

\newcommand{\id}{\mathrm d}
\newcommand{\vc}{\boldsymbol}

\newcommand{\gr}{\mbox{g}}

\renewcommand{\tilde}{\widetilde}
\newcommand{\pard}[2]{\frac{\partial #1}{\partial #2}}

\newcommand{\longcomment}[1]{}

\newcommand{\grav}{\text{g}}
\newcommand{\optcoeff}{\tilde{c}}

\newcommand{\sensor}{s_1}

\begin{document}
\title{Reconstruction and Forecast of Extreme Waves from a Single Observation}

\author[1]{Zachary T. Hilliard} 
\affil[1]{Department of Mathematics, Tufts University, 177 College Avenue, Medford, Massachusetts 02155, USA} 

\author[2,3,4]{Amin Chabchoub}
\affil[2]{Marine Physics and Engineering Unit, Okinawa Institute of Science and Technology, Onna-son, Okinawa 904-0495, Japan} 
\affil[3]{Department of Civil and Environmental Engineering, Imperial College London, London, SW7 2AZ, United Kingdom}
\affil[4]{Department of Infrastructure Engineering, The University of Melbourne, Parkville, Victoria 3010, Australia }

\author[5]{Mohammad Farazmand\thanks{Corresponding author: farazmand@ncsu.edu}}
\affil[5]{Department of Mathematics, North Carolina State University, 2311 Stinson Drive, Raleigh, North Carolina 27695, USA} 

\date{}
\maketitle

\begin{abstract}
The state space of a dynamical system can, under suitable conditions, be reconstructed from a single observational time-series through delay-coordinate embedding. However, reconstructing the spatiotemporal evolution of nonlinear water waves from highly sparse measurements remains a considerable challenge. Here, by combining deep learning, compressed sensing, and the self-similarity properties of unidirectional nonlinear water wave evolution equation, we show that the evolving surface elevation of a highly nonlinear wave field can be reconstructed from observations collected at a {\it single} measurement location. We subsequently use these reconstructions to forecast rogue waves associated with self-focusing soliton dynamics from point observations obtained in controlled water wave tank experiments, demonstrating excellent agreement with measurements of the fully evolved wave field. Our results show that extreme wave events can be forecast from remarkably sparse observations and provide a framework that may be extended to other nonlinear dispersive systems governed by the interplay of dispersion and nonlinearity.
\end{abstract}

\section{Introduction}
Rogue waves, also referred to as freak waves or killer waves, are short-lived waves of exceptionally large amplitude that have been observed in a variety of nonlinear dispersive media, including hydrodynamic, optical, plasma, and Bose-Einstein condensate systems~\cite{haver2004possible,Solli2007,kharif09,Onorato13,dudley2019rogue,mori2023science,chabchoub2026extreme}. With extreme wave heights reaching up to 30 meters, oceanic rogue waves naturally pose a significant risk to marine vessels and offshore structures, such as oil platforms and marine observation facilities~\cite{Nikolkina2011,Didenkulova2023}. As such, the early prediction of their formation is of considerable interest across disciplines.

It is now well-established that extreme waves can arise either from wave interference~\cite{mcallister2024three,Knobler2025} or from nonlinear interactions associated with four-wave resonant interactions and modulation instability (MI)~\cite{benjamin67,Zakharov2009}. Such nonlinear mechanisms have been confirmed in the ocean~\cite{maestrini2026revealing} and have motivated a number of controlled laboratory studies in water wave facilities~\cite{onorato2006extreme,Chabchoub2011,houtani2018experimental}. In a simplified framework, rogue waves that emerge from modulationally unstable backgrounds can be deterministically modeled using breather solutions of the nonlinear Schrödinger equation (NLS)~\cite{Zakharov68,akhmediev1997solitons}, enabling their controlled generation in a variety of nonlinear wave systems~\cite{Kibler2010,Chabchoub2011,bailung2011observation,romero2024experimental}. Indeed, such breathers can also be understood as solitonic or multi-solitonic perturbations of a regular background~\cite{chabchoub2021peregrine}. 

What makes rogue waves particularly dangerous is their emergence with no readily observable precursor, rendering their prediction particularly difficult. Nonetheless, recent studies have suggested that MI only amplifies wave packets that satisfy a particular combination of amplitude, length scale, and phase~\cite{cousins15,cousins16}. This observation has enabled the forecasting of rogue waves with lead times of several wave periods~\cite{Farazmand2017,PRE2016,Dematteis2018,Breunung2024}. However, identifying such precursors requires measurements of the wave elevation over a dense spatial grid. More recently, significant theoretical and applied progress has been made toward predicting nonlinear wave interactions using data-driven and related computational approaches~\cite{eeltink2022nonlinear,adriazola2025machine,boussafa2025deep,kavousanakis2026going}.

In this paper, we show that unidirectional rogue waves, arising from multi-soliton dynamics~\cite{Satsuma1974}, can be predicted from a time series of wave elevation measured at a single stationary point. Takens' delay-embedding theorem establishes that the state of a dissipative dynamical system can, under suitable conditions, be reconstructed from a single time series~\cite{takens,sauer91}. However, water waves constitute a conservative system, making direct reconstruction of the surrounding wave field from a single-point measurement challenging. Here, through a synergistic combination of deep learning, compressed sensing, and scaling symmetries, we show that experimentally measured wave elevation at a single point can be used to infer the wave field in the neighborhood of the wave gauge. Subsequently, a wave-propagation NLS model is used to forecast upcoming rogue waves up to 38 wave periods in advance. Our predictions reproduce the observed rogue-wave amplitude with an error of approximately 8\%, while the occurrence time is predicted with an accuracy of approximately 0.5\%.

\section{Experimental set-up}
\begin{figure}[t!]
\centering
\includegraphics[width=.8\columnwidth]{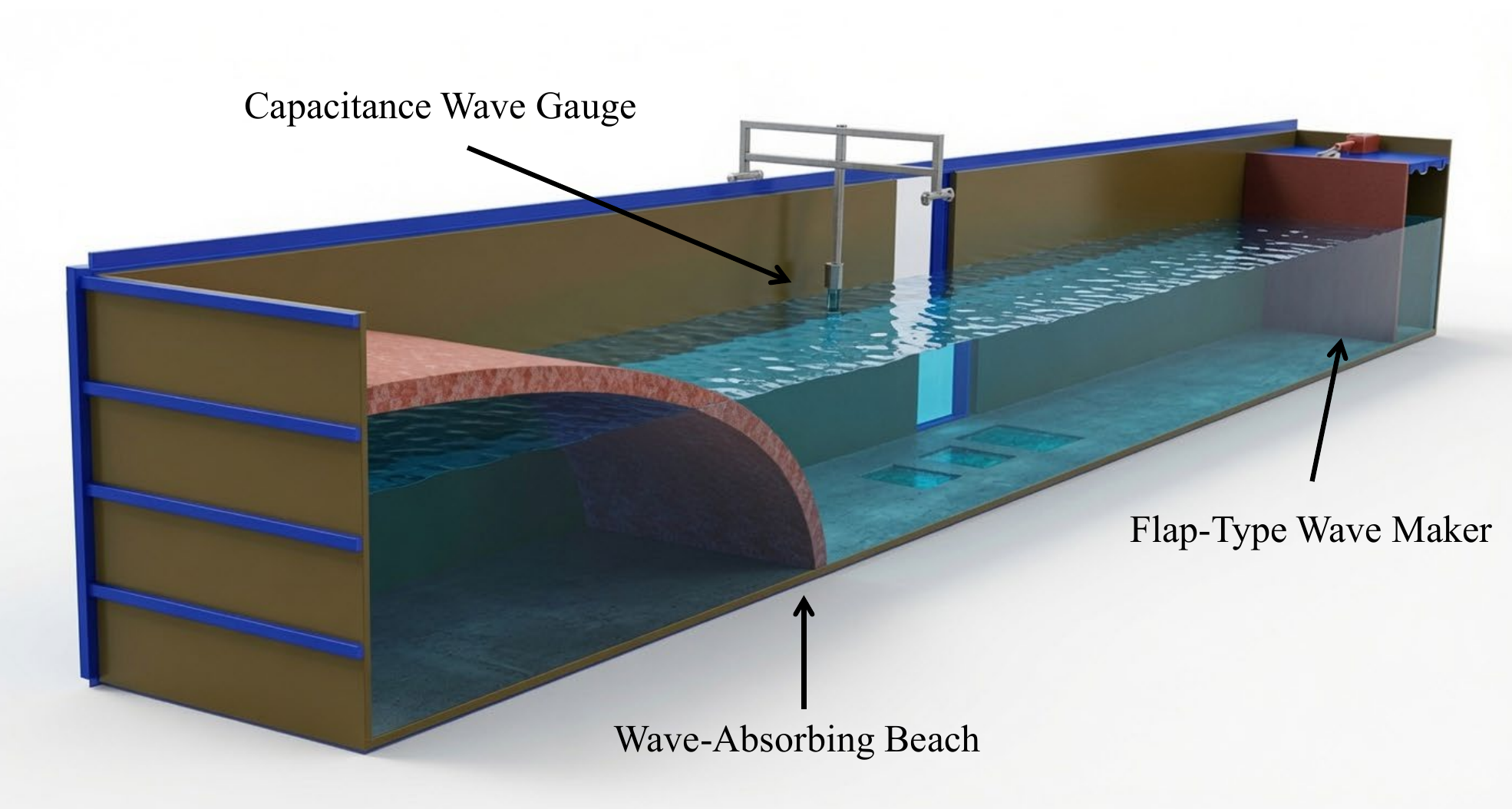}
\caption{Schematic of the water wave tank apparatus, showing the flap-type wave maker, movable capacitance wave gauge, and wave-absorbing beach installed at the opposite end of the tank.}
\label{fig:exp_setup}
\end{figure}

We begin by describing the experimental data for the large-amplitude hydrodynamic multi-soliton wave packet generated in a controlled laboratory setting. The experiments were conducted in a wave tank 15~m long, 1.6~m wide, and 1.5~m deep, with an effective water depth of 1~m. A flap-type wave maker located at one end of the tank generates a second-order soliton, or two-soliton solution, with an initial carrier amplitude $a_0=0.005$~m and wavelength $\lambda_0=0.39$~m, corresponding to the angular frequency $\omega_0=\sqrt{\gr k_0}=12.5$ $\operatorname{s}^{-1}$, where $k_0=2\pi/\lambda_0=16$ $\operatorname{m}^{-1}$ denotes the carrier wavenumber and $\gr$ is the gravitational acceleration. We refer to Ref.~\cite{Chabchoub2013a} for the exact parametrization of the second-order soliton and the specification of the boundary condition. A synthetic beach is installed at the opposite end of the tank to minimize wave reflections. Wave gauges are positioned along the tank at intervals of $0.2$~m and measured the surface elevation $\eta(s_i,t)$ at a sampling frequency of 100~Hz. For the present analysis, we consider only a subset of these measurements, corresponding to gauge locations $s_i=s_1+(i-1)\Delta x$, where $s_1=1.8$~m is located close to the wave maker, $\Delta x=1$~m, and $i=1,\ldots,9$. A schematic of the experimental setup is shown in Fig.~\ref{fig:exp_setup}. Here, we use only the data recorded by the gauge closest to the wave maker, located at $x=s_1=1.8$~m, to infer the downstream wave elevation $\eta(s_i,t)$ along the tank. Measurements from the remaining eight locations are used exclusively to validate the reconstruction and quantify the accuracy of the predictions.

\section{Symmetries, compressed sensing and training}
\begin{figure*}
	\centering
	\includegraphics[width =  \textwidth]{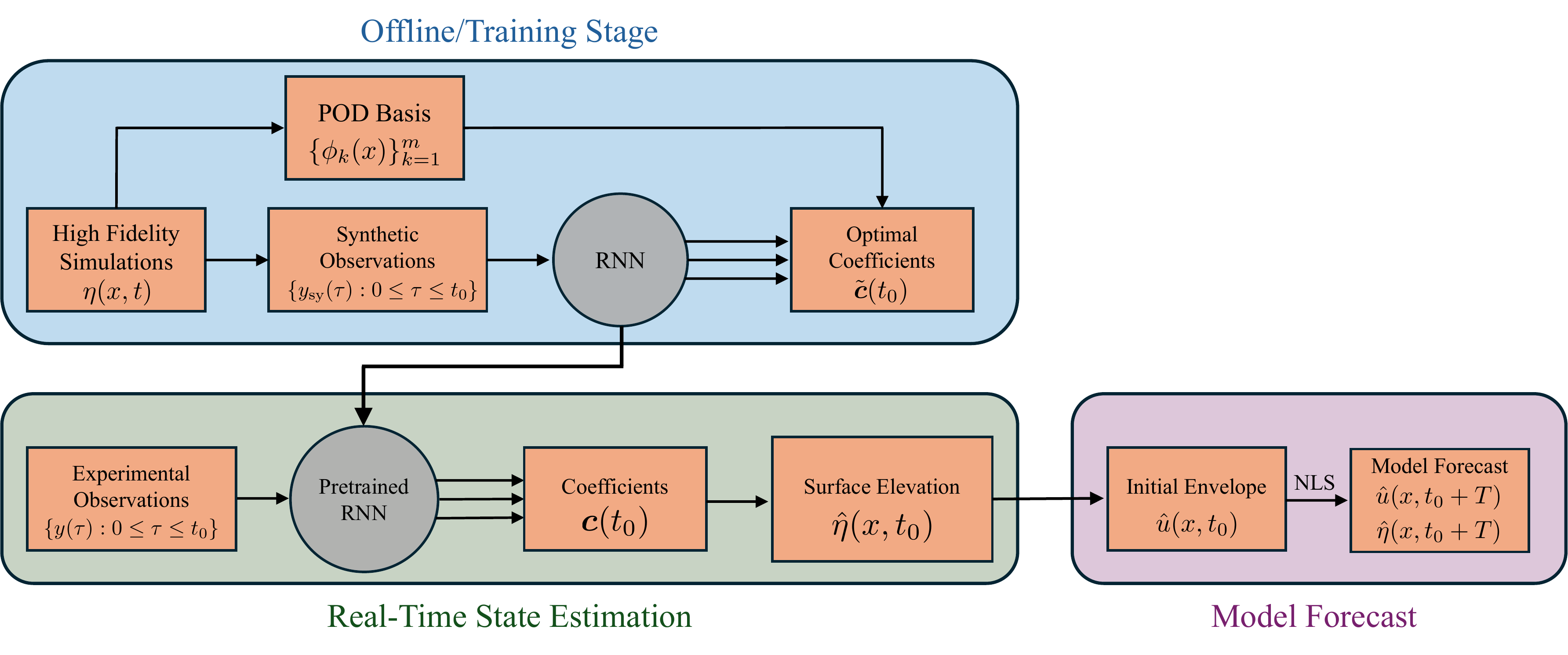}
	\caption{Illustration of the workflow. In the top panel, offline/training stage, we use sythetic data from NLS to train the recurrent neural network (RNN). The pretrained RNN is then used in the bottom-left panel to reconstruct the surface elevation from experimental data. Using the Hilbert transform and phase retrieval, the surface elevation is tranformed into the wave envelope which is subsequently used as the initial condition for NLS to forecast the dynamics.}
	\label{fig:flow_chart}
\end{figure*}

Our predictions rely on the inference of the wave elevation $\eta(x,t_0)$ from the pointwise time series $y(\tau) = \eta(s_1,\tau)$, where $0\leq \tau\leq t_0$ 
To this end, we train a recurrent neural network (RNN) which takes the observational time series $y(\tau)$ as input. The training process requires a large amount of data which is far beyond the available experimental data described above. Furthermore, since rogue waves are rare, only a small portion of long-term experiments would correspond to these extreme waves. This is a well-known bottleneck for training deep neural networks for rare events \cite{farazmand2019a, Asch2022, Champenois2025}. To address this issue, we train our RNN using synthetic data which mimic the wave tank experiments. As depicted in Fig.~\ref{fig:flow_chart}, the trained neural networks is subsequently used to predict rogue waves from experimental data. The synthetic data are generated from the NLS equation,
\begin{equation}\label{eq:dim_NLS}
\hat{i}\left(\pard{u}{t} + c_g \pard{u}{x}\right)-\frac{\omega_0}{8 k_0^2}\pard{^2u}{x^2} - \frac{\omega_0k_0^2}{2}|u|^2u = 0,
\end{equation}
which models slowly modulating waves in deep-water~\cite{Zakharov68}. 
Here, $\hat i=\sqrt{-1}$ and $u(x,t)$ denotes the complex-valued wave envelope. The independent variables are the spatial coordinate $x$ and time $t$. The constants $k_0$ and $\omega_0$ denote the wave number and angular frequency of the carrier wave, respectively. The group velocity $c_g$ is given by $c_g = \frac{\partial \omega }{\partial k}|_{k = k_0}=  \frac{\omega_0}{2k_0}$. The motivation for using the NLS equation, instead of the full water-wave equations, is twofold. First, the NLS does not require knowledge of the velocity potential which is challenging to measure in experiments. More importantly, the NLS admits exact multi-soliton and breather solutions which describe rogue wave dynamics \cite{akhmediev1997solitons}, which have been reproduced experimentally in hydrodynamics and other wave systems~\cite{Mollenauer1980,Kibler2010,Chabchoub2011,bailung2011observation,Chabchoub2013a,romero2024experimental}. 

Given a solution of NLS, the surface elevation is given by
\begin{equation}\label{eq:eta}
\eta(x,t) = \text{Re}\left[u(x,t) \exp \left(\hat{i}(k_0 x - \omega_0 t)\right)\right]. 
\end{equation}
If the initial envelope $u_0(x) := u(x,0)$ is known, the NLS equation can be used to forecast the envelope $u(x,t)$ and subsequently compute the surface elevation $\eta(x,t)$ for any $t\geq 0$. However, as mentioned earlier, neither the initial envelope $u_0(x)$ nor the corresponding surface elevation $\eta_0(x)$ are known; only the surface elevation $y(\tau):=\eta(s_1,\tau)$ at a fixed sensor location $x=s_1$ is known. 

Instead, we train an RNN to reconstruct the surface elevation $\eta(x,t_0)$ given the observations $y(\tau)$ for $\tau\leq t_0$. The synthetic training data is obtained from 121 solutions of the NLS Eq.~\eqref{eq:dim_NLS}, each corresponding to a different initial condition $u_0(X) = 2\alpha \text{sech}(\alpha X)$ where $X=\sqrt{2}k_0x$ is the dimensionless spatial variable, and the amplitude $\alpha$ correspond to 121 equidistant points within the interval $[0.04,0.1]$. The shape of the initial conditions correspond to self-similar 2-soliton solutions of NLS which are known to mimic rogue waves \cite{Chabchoub2013a,Satsuma1974}. 
We numerically integrate the NLS equation, using a Fourier pseudo-spectral method with an exponential time-differencing scheme, and compute the corresponding surface elevations using Eq.~\eqref{eq:eta}.
Further details regarding the numerical solver are provided in Appendix~\ref{app:numerics}.

The input of the RNN is the single time series $y(\tau)$. In principle, the outputs can comprise the surface elevation $\eta(x,t_0)$ on a sufficiently dense spatial grid. However, the number of outputs will be so large, compared to the single input, that the RNN will struggle to accurately predict the output. Instead, we take a compressed sensing approach, using Proper Orthogonal Decomposition (POD)~\cite{berkooz93,Shlizerman2012}, where the surface elevation $\eta(x,t)$ is approximated by 
\begin{equation}\label{eq:POD_recon}
\hat{ \eta}(x,t) = \sum_{k=1}^m c_k(t) \phi_k(x), 
\end{equation}
where $\{\phi_k (x)\}_{k=1}^m$ are the POD modes and $c_k(t)$ are the corresponding POD coefficients. The POD modes are extracted from the same 121 synthetic training data obtained from solving the NLS.

It is well-known that such advection dominated systems will require a large number of POD modes in order to achieve an accurate approximation~\cite{Hesthaven2022,Farazmand2026b}. More precisely, according to the Kolmogorov $m$-width theory, the approximation error decays only with a rate of $m^{-1/2}$, instead of exponentially~\cite{Ehrlacher2020}. We overcome this issue by transferring the NLS solutions to a co-moving frame that travels with the group velocity $c_g$. In the co-moving frame, the waves appear almost stationary and therefore far fewer POD modes are needed to achieve a desired approximation error threshold. In particular, in the co-moving frame, $m=15$ POD modes are sufficient to ensure that $\hat\eta$ approximates the true wave elevation $\eta$ with a relative error of less than 0.05\%, if $\eta(x,t)$ is completely known.

At any time $t$, the optimal set of coefficients $\optcoeff_k$ is given by $\optcoeff_k(t) =\left\langle\phi_k, \eta(\cdot,t)\right\rangle_{L^2}$, where $\eta(\cdot,t)$ is the synthetic surface elevation in the co-moving frame. During training, these coefficients constitute the target outputs. The input of the network is the time series $y(\tau) = \eta(s_1(\tau),\tau)$, $\tau\leq t_0$, evaluated at the sensor location $s_1(\tau)$. Note that the stationary sensor appears non-stationary in the co-moving frame, with its location given by $s_1(\tau)=s_1-c_g \tau$. To mimic the experimental data, we set $s_1=1.81$\,m. As such, there might be times $\tau$ at which the location of the sensor $s_1(\tau)$ does not precisely coincide with any point in the computational grid. In such cases, we
evaluate $y(\tau)$ by linear interpolation between the two closest grid points that contain $\sensor(\tau)$.
\begin{figure}[t!]
\centering
\includegraphics[width = \columnwidth]{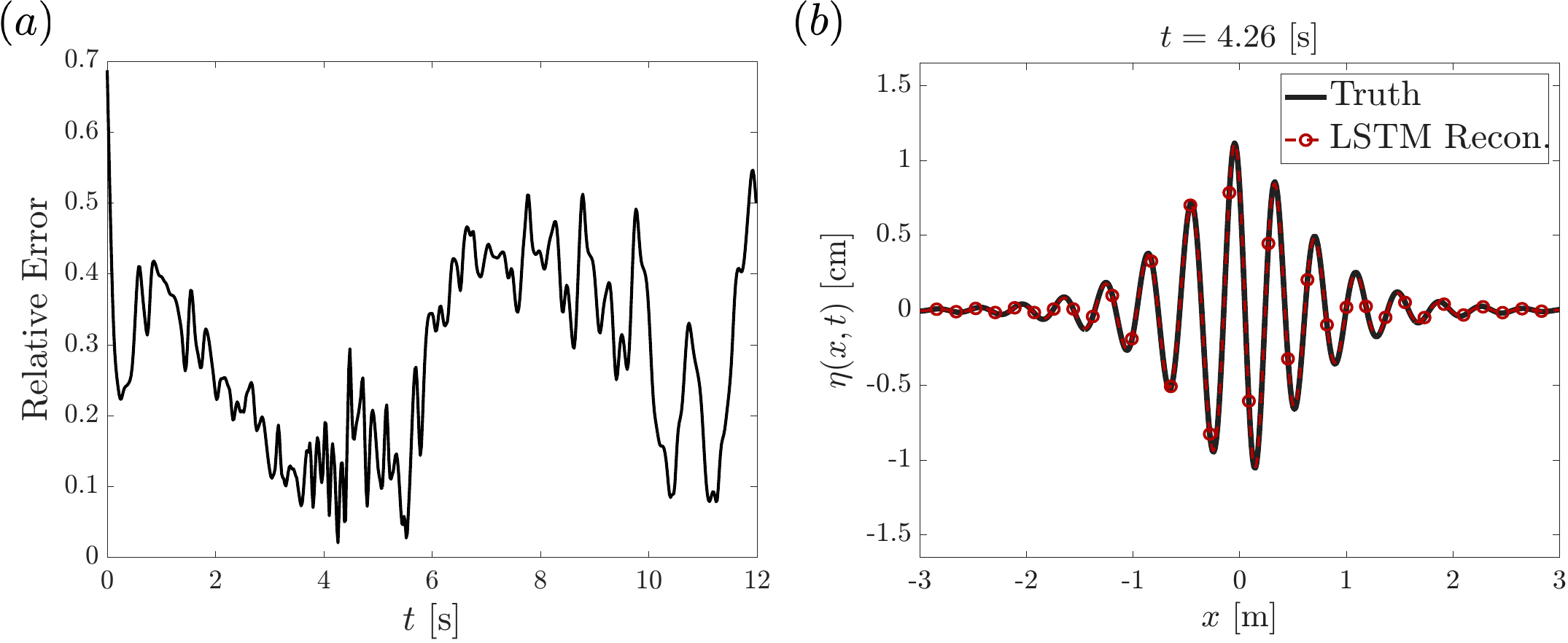}
\caption{LSTM reconstruction of the wave elevation on the synthetic data. (a) The relative error between the LSTM estimate and the true surface elevation. (b) The LSTM reconstruction of the surface elevation at $t = 4.26$\,s (red dashed line) on top of the NLS solution (solid black line). The relative reconstruction error is 2.1\%.}
\label{fig:synthetic} 
\end{figure}

We note that compressed sensing methods, such as Discrete Empirical Interpolation Method (DEIM)~\cite{barrault2004,Sorensen2010}, usually infer the POD coefficients $c_k(t)$ from the observations $y(t)$ using closed-form expressions~\cite{Eldar2012}. However, these expressions do not return satisfactory reconstructions when the observations are extremely sparse~\cite{Farazmand2024a,Farazmand2024b,All2026}; in fact, only one observation is available in our case. This motivated our use of RNNs as a black-box, yet powerful, alternative.

Using the 121 solutions of NLS, we train an RNN with input $\{y(\tau), 0\leq\tau\leq t_0\}$ and outputs $\tilde{\vc c}(t_0)$. 
The outputs are translated into the surface elevation through Eq.~\eqref{eq:POD_recon}.
For the RNN, we choose a long short-term memory (LSTM) network, which is effective at predicting time series data by retaining information about the past history of the time series~\cite{Sepp1997, Asch2022}. 

The RNN has 100 hidden LSTM units followed with a two-layer fully connected regression unit which maps the LSTM features to the outputs. We train the network using the Adam optimizer~\cite{kingma2017} over 1500 epochs. Further details regarding the training process and the network architecture are given in Appendix~\ref{app:rnn}. 

\section{Results and discussion}
\subsection{Synthetic data}
Figure~\ref{fig:synthetic} shows the results for a representative sample of the synthetic training data. As shown in Fig.~\ref{fig:synthetic}(a), the relative error, $\|\hat\eta-\eta\|_{L^2}/\|\eta\|_{L^2}$, is initially large at around 70\% because the wave packet has not yet reached the wave gauge and the observations are not yet informative. However, the error decays rapidly until it reaches its minimum of 2.1\% at time $t_0=4.26$\,s. This time corresponds approximately to the time at which the peak of the wave group reaches the sensor. This trend is observed across the training data, and therefore we choose the reconstruction time $t_0$ to coincide with the time at which the maximum wave envelope is registered by the sensor. Note that the reconstruction time $t_0$ significantly precedes the formation of the rogue wave which occurs around $t=18$\,s. Figure~\ref{fig:synthetic}(b) shows the LSTM reconstruction of the wave elevation $\hat\eta(x,t_0)$ which is in excellent agreement with the true wave elevation. The reconstruction error eventually increases again, as the wave packet travels away from the gauge.

\subsection{Experimental data}
Once the RNN is trained on synthetic data, it is used to infer the surface elevation in experiments, as depicted by the green box in Fig.~\ref{fig:flow_chart}. In this stage, we use the experimental observations $\{y(\tau)=\eta(s_1,\tau): 0\leq \tau\leq t_0 \}$ gathered at the first sensor location $\sensor=1.80$\,m as inputs to the pre-trained RNN. The network outputs predict the POD coefficients $\vc c(t_0)$ which are used in~\eqref{eq:POD_recon} to estimate the surface elevation $\hat \eta(x,t_0)$. 

To proceed to the forecasting stage, depicted by the purple box in Fig.~\ref{fig:flow_chart}, we need to infer the wave envelope $\hat{u}(x,t_0)$ from the estimated wave elevation $\hat{\eta}(x,t_0)$. To this end, we first use the Hilbert transform to compute the complex modulus $|\hat u(x,t_0)|$ of the wave envelope $\hat u(x,t_0) = |\hat u(x,t_0)| \exp[\hat i \varphi(x,t_0)]$. However, the phase $\varphi(x,t_0)$ is not correctly identified by the Hilbert transform. To approximate the correct phase, we discretize it using Fourier modes and use a standard optimization routine to solve for the appropriate Fourier coefficients of the phase. Although the phase does not significantly affect the amplitude of the ensuing rogue wave dynamics, it can impact the time, and therefore the location, at which the rogue wave forms~\cite{Farazmand2017,chabchoub2020phase}. We refer to the Appendix~\ref{app:env} for the details of the Hilbert transform and the phase retrieval method.
\begin{figure}[t!]
\centering
\includegraphics[width =  \columnwidth]{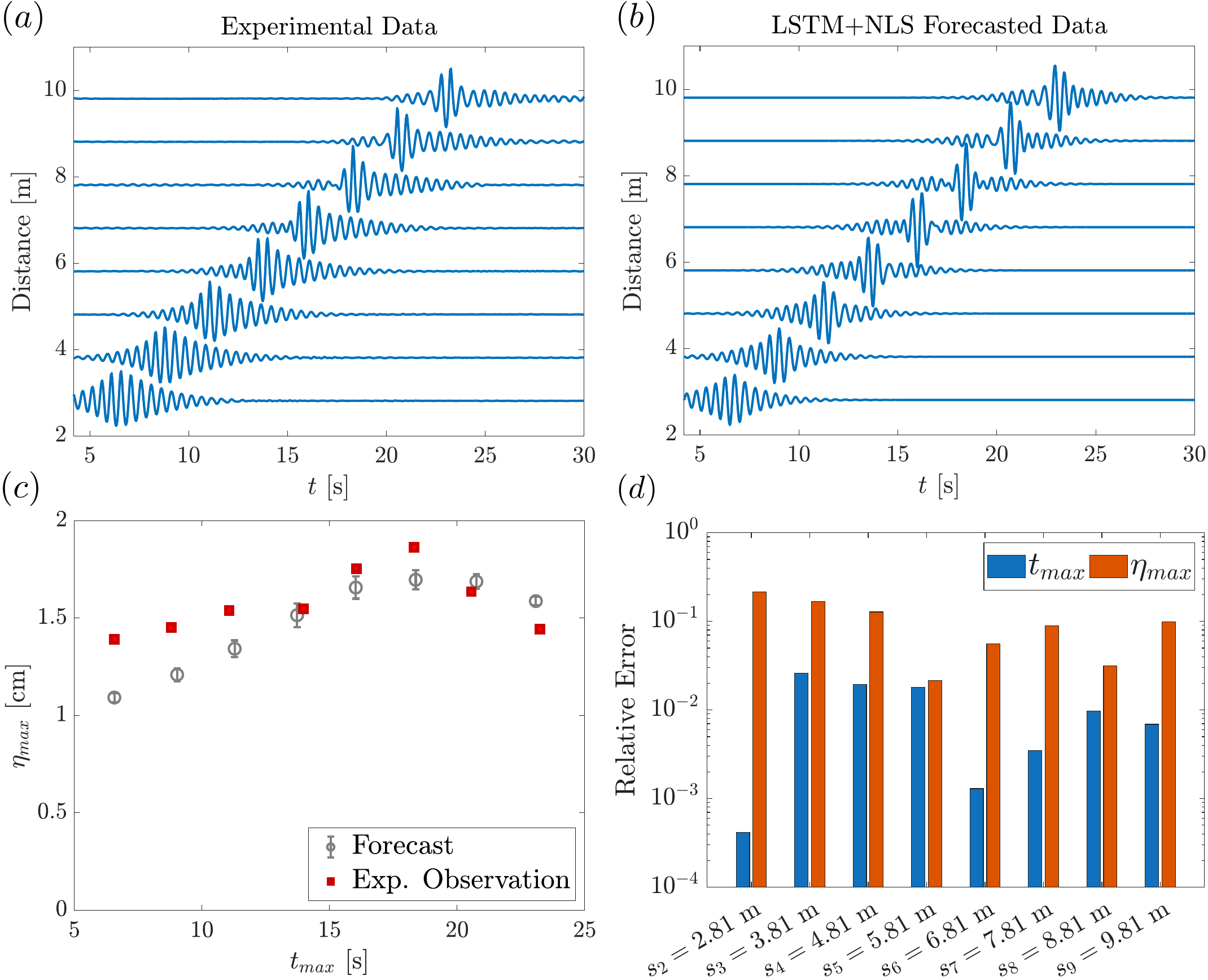}
\caption{Comparing forecasts with experiments. (a) the experimental data at eight remaining sensor locations, spaced 1 meter apart. (b) The  predictions using the LSTM reconstruction and the NLS forecast. (c) The average value of the maximum surface elevation at each sensor location with the corresponding time at which the maximum occurs. The error bars mark the 90\% confidence intervals around the mean. (d) The relative errors of the time when the maximum occurs (blue bars on the left) and the maximum surface elevation (orange bars on the right).}
\label{fig:data_comp}
\end{figure}

We use the estimated wave envelope $\hat u(x,t_0)$ as the initial condition for the NLS Eq.~\eqref{eq:dim_NLS} to forecast the wave evolution for $t\geq t_0$ as it travels downstream in the wave tank.
We compare our forecasts to true experimental observations in Fig.~\ref{fig:data_comp}. Panels (a) and (b) show the time series of the experimental observations and the corresponding NLS forecasts at eight remaining sensor locations. Qualitatively, we see good agreement between the experimental results and the forecasts. 
The scatter plot in Fig.~\ref{fig:data_comp}(c) shows the maximum amplitude, at each sensor location, predicted by NLS and the time at which the maximum occurs; these are compared with the corresponding experimental measurements. The error bars of NLS forecasts correspond to 90\% confidence intervals by varying the reconstruction time $t_0$ inside the interval $[2,6]$ seconds, showing robustness to the reconstruction time. We see that both the maximum amplitudes and the occurrence times agree well with the experiments. More specifically, Fig.~\ref{fig:data_comp}(d) shows the relative errors of the maximum wave amplitude and the occurrence time. The rogue wave reaches its peak at around $t=18$\,s, as registered by the 7th sensor $s_7$. At this location, the NLS forecast exhibits an approximately 8\% error in the amplitude as compared with the true value from experiments. Moreover, the timing of the formation of the rogue wave is predicted even more accurately with only 0.5\% error. 

\section{Conclusions}
In conclusion, we have demonstrated that extreme wave elevations in a wave tank experiment can be reconstructed and forecast from measurements at a single upstream sensor. Our approach combines compressed sensing in a POD basis with an RNN trained on synthetic NLS data. Transforming the wave fields into a co-moving frame mitigates the slow decay of the Kolmogorov $m$-width, while the self-similarity of NLS solutions enables the construction of training data representative of the experimentally observed two-soliton dynamics. Applied to experimental measurements, the resulting framework successfully forecasts an upcoming rogue wave $14~\mathrm{s}$ in advance, corresponding to approximately 38 wave periods.

Minor discrepancies between the forecasts and the experimental observations arise primarily from higher-order effects neglected by the NLS equation, including amplitude-dependent corrections to the group velocity and asymmetry of the wave envelope and mean flow effects. Such dynamics can be captured by modified NLS type models~\cite{dysthe79,Chabchoub2013a,Dudley2006} and will be considered in future work. While the present experiments are restricted to unidirectional, narrowband waves, extending the framework to broadband and directionally spread ocean waves constitutes an important next step. More broadly, the proposed methods may be applicable to other nonlinear dispersive systems in which extreme wave events emerge as a consequence of the interplay of dispersion and nonlinearity. 

\textbf{Acknowledgments}---M.F. and Z.T.H. were partially supported by the U.S. National Science Foundation awards, DMS-1745654 (Research Training Group), DMS-2208541 (Computational Mathematics Program), and DMS-2220548 (Algorithms for Threat Detection). A.C. acknowledges support from
Okinawa Institute of Science and Technology (OIST)
with subsidy funding from the Cabinet Office, Government of Japan.

\appendix

\section{Computation of POD Modes}\label{app:POD}
To obtain the POD modes, we must first solve the nonlinear Schr\"odinger (NLS) equation with a variety of initial conditions. To do so, we first non-dimensionalize NLS by introducing the following non-dimensional variables: 
\begin{equation}
X =  \sqrt{2} k_0 x , \quad T = \frac{\omega_0}{4} t,\quad  \text{and} \quad U = k_0 u,
\end{equation}
where $k_0 = 16$ is the wave-number and $\omega_0 = \sqrt{\grav k_0}$ is the frequency of the carrier wave and $\grav = 9.81 \ \text{m}/\text{s}^2$. Lastly, we know that solutions to NLS propagate at a fixed speed, namely the group velocity $c_g$, we transform to a co-moving frame with speed $c_g$. Under these transformations, the non-dimensional NLS reads,
\begin{equation}
\hat{i}U_T + U_{XX} + 2U|U|^2 = 0, 
\end{equation}
with initial data $U(X,0) = U_0(X)$. The associated boundary conditions for NLS are that the solution vanishes as $|X|$ approaches infinity. 

To compute the POD modes we evolve the initial conditions that correspond to the 2-soliton solution~\cite{Chabchoub2013a,Satsuma1974}, 
\begin{equation}\label{eq:ICs}
u_i(X) = 2 \alpha_i \text{sech}(\alpha_i X)
\end{equation}
where $\alpha$ comes from the self-similarity of solutions to NLS~\cite{Chabchoub2016a}. In order to ensure that the trajectories for POD adequately capture the dynamics of the wave-tank, we choose 121 values of $\alpha_i$ in the interval $[0.04, 0.1]$. Namely, we take 
 \begin{equation}
\alpha_i  = 0.04 + 5\times 10^{-4}(i-1), \quad \text{for} \quad i = 1,2,\ldots, 121. 
 \end{equation} 
 In addition, these values of $\alpha_i$ ensure that the solutions are away from the wave-breaking limit~\cite{Chabchoub2013a}, so that they are physically reasonable.
 
\section{Numerical Integration} \label{app:numerics}
We approximate NLS using a Fourier psuedo-spectral truncation in space with $2^{12}$ Fourier modes, which corresponds to an equivalent number of grid points in the spatial domain $\Omega$. To approximate the infinite boundary conditions of the PDE, we let $\Omega = [-25, 25]$~m and use periodic boundary conditions. We integrate over the time interval $[0,35]$~s with an exponential time-differencing scheme as described in Ref.~\cite{Cox2002} and a time step of $\Delta t = 1/50$ s. To handle the nonlinearity, we use a standard 2/3rds dealiasing routine. 

\begin{figure}[!h]
\centering 
\includegraphics[width = \columnwidth]{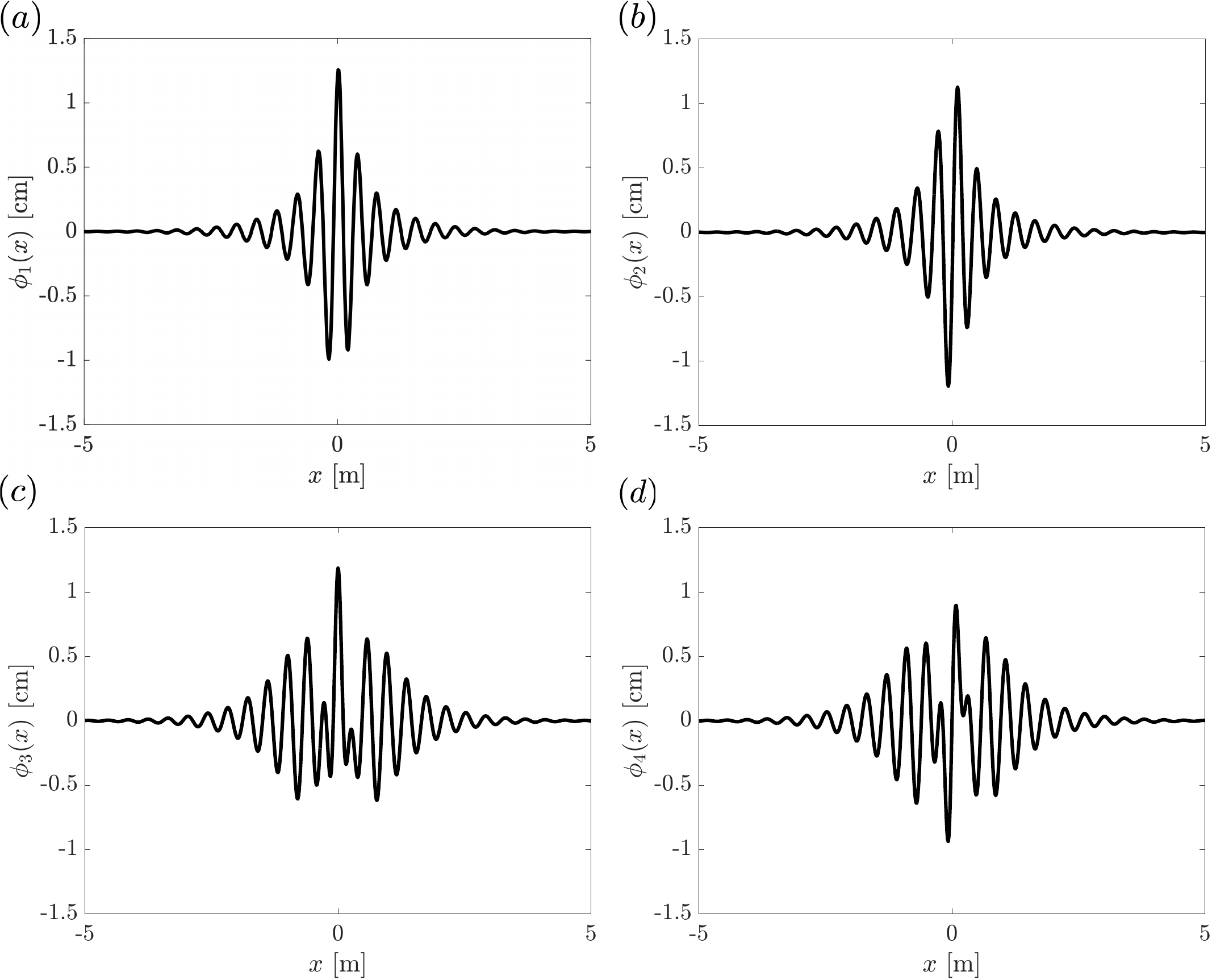}
\caption{\label{fig:POD_modes}First four POD modes for the data corresponding to the breather 2-soliton solution of NLS.}
\end{figure}

Next, using the relationship between the surface elevation $\eta$ and the envelope $u$, 
\begin{equation}\label{eq:u2eta}
	\eta(x,t) = \text{Re}\left[u(x,t) \exp \left(\hat{i}(k_0 x - \omega_0 t)\right)\right],
\end{equation}
we obtain the corresponding surface elevations. We use these values of $\eta$ to compute a set of $m$ spatially dependent POD modes $\{\phi_k (x)\}_{k=1}^m$. We include the first four POD modes in Fig.~\ref{fig:POD_modes}.
  
\section{Recurrent Neural Network: Architecture and Training Parameters}\label{app:rnn}
\begin{figure}[htbp]
\includegraphics[width = \columnwidth]{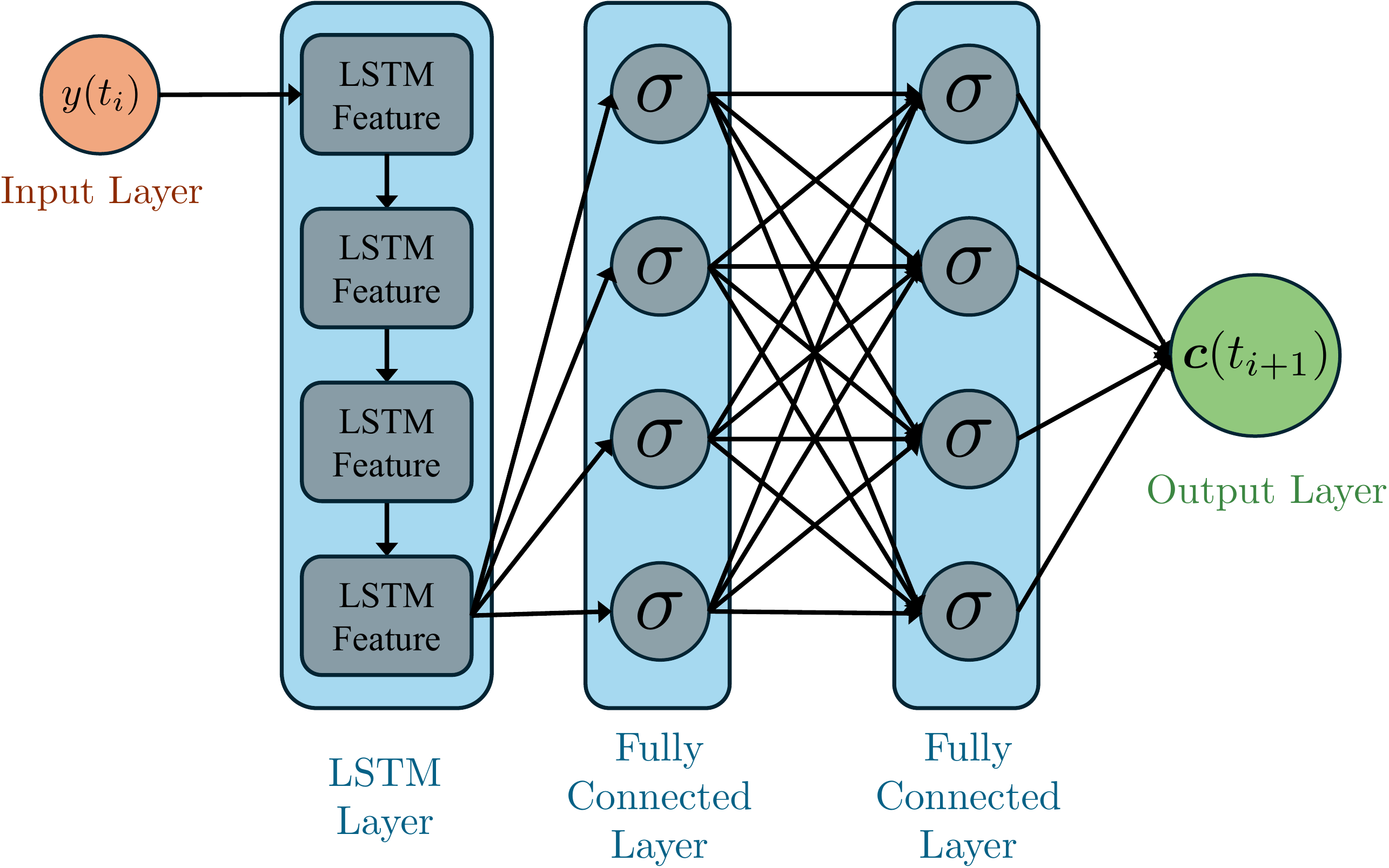}
\caption{\label{fig:architecture}Recurrent neural network architecture.}
\end{figure}
To build and train the recurrent neural network (RNN) we use MatLab's Deep Learning Toolbox. For our study, the RNN has 1 input layer and 1 long-short term memory (LSTM) layer with 100 hidden units. These layers are followed by 2 fully connected layers with 15 channels each which corresponds to the number of POD modes used in the state estimate. We provide an illustration of our chosen architecture in Fig.~\ref{fig:architecture}. To train the LSTM, we use Matlab's built in Adam~\cite{kingma2017} optimization routine. We train for a maximum of 1500 epochs, use an L2 regularization of $10^{-8}$, and a learning rate of 0.001. In addition, we specify the root mean square error as our loss function for the optimizer.

\section{Complex-Valued Envelope Reconstructions}\label{app:env}
The RNN reconstructs the surface elevation $\hat\eta(x,t)$ from sparse observations. However, in order to forecast with NLS, we require the wave envelope $\hat u(x,t)$. To reconstruct the envelope, we use the Hilbert transform $\mathcal H$ which is defined as the convolution of the wave elevation with the impulse response function $\frac{1}{\pi x}$,
\begin{equation}
\mathcal{H}(\hat\eta) = \frac{1}{\pi} \ \text{p.v.} \ \int_{-\infty}^{\infty} \frac{\hat\eta(\xi,t)}{x-\xi}\id \xi,
\end{equation}
where $\text{p.v}$ denotes the Cauchy principle value to handle the singularity at the origin~\cite{King2009}.

The so-called analytic signal corresponding to $\hat\eta$ is given by the complex-valued function,
\begin{equation}
	\hat\eta_a(x,t) = \hat\eta(x,t) + \hat{i} \mathcal{H}(\hat \eta).
\end{equation}
MatLab's command \texttt{hilbert($\eta$)} computes this analytic signal.
It is known that $|\hat\eta_a|$ coincides with the envelope of $\hat \eta$~\cite{Marple1999}, i.e., $|\hat u(x,t)|=|\hat\eta_a(x,t)|$. 
Although the modulus of the analytic signal returns the correct envelope, the phase $\varphi(x,t)$, where $\hat u(x,t) = |\hat\eta_a(x,t)| \exp(\hat i\varphi(x,t))$, may not be correct.

One way to find the correct phase is to assume that $\varphi$ is constant in space, and solve the following optimization problem: 
\begin{equation}\label{eq:phase_opt_const}
\min_{\varphi \in [0,2\pi]} \frac{1}{2}\left\|\hat{\eta} - \text{Re}\left[|\hat{\eta}_a|\exp\left(\hat{i}(k_0x-\omega_0 t+\varphi)\right)\right]\right\|_{L^2}^2,
\end{equation}
where we used Eq.~\eqref{eq:u2eta}. This problem can be solved by simply discretizing the interval $[0,2\pi]$ and finding the value of $\varphi$ which corresponds to the minimizer of the functional defined by the norm in Eq.~\eqref{eq:phase_opt_const}. In practice, however, these envelope reconstructions led to inaccurate results when forecasting with NLS, indicating that the constant-phase assumption is too restrictive.

A better assumption is to allow the phase $\varphi$ to be spatially dependent and instead solve,
\begin{equation}
\min_{\varphi\in\mathcal L^2}\frac{1}{2}\left\|\hat{\eta} - \text{Re}\left[|\hat{\eta}_a|\exp\left(\hat{i}(k_0x-\omega_0 t+\varphi(x))\right)\right]\right\|_{L^2}^2. 
\end{equation}
We discretize this problem using the Fourier transform of the phase,
\begin{equation}
\varphi(x;\vc a) = \sum_{j} a_j(t) \exp\left[{\hat{i} \frac{2\pi j}{L} x}\right] + c.c.,
\end{equation}
where $L$ is the length of the spatial domain $\Omega$. We then optimize over the Fourier coefficients $\vc a = [\cdots,a_{-1},a_0,a_1,\cdots]$. Since we approximate the infinite boundary conditions of NLS with a large spatial domain and periodic boundary conditions, the Fourier modes are a good choice of basis function for this approximation.
Truncating the Fourier series at $N$ modes, we solve the finite-dimensional optimization problem,
\begin{equation}\label{eq:phase_opt}
\min_{\vc a\in\mathbb C^N} \frac{1}{2}\left\|\hat{\eta} - \text{Re}\left[|\hat{\eta}_a|\exp\left(\hat{i}(k_0x-\omega_0 t+\varphi(x;\vc a))\right)\right]\right\|_{L^2}^2,
\end{equation}
using standard optimization routines. For our study, we use MatLab's \texttt{fminunc}, where the relations $a_{-j}=a_j^\ast$ are used as constraints, ensuring that the phase is real-valued. To supply an appropriate initial guess for optimization, we take the constant solution from Eq.~\eqref{eq:phase_opt_const} and use that as the initial value of the parameters $\vc a$. In line with the discretization of NLS, we dealiase the vector $\vc a$ using a standard 2/3 dealiasing routine. Lastly, we reconstruct the approximate envelope $\hat{u}$ 
 \begin{equation}
 \hat{u}(x,t) = \left|\hat\eta_a\right|\exp\left(\hat{i}(k_0x-\omega_0t+\varphi(x; \vc a))\right).
 \end{equation}
  \begin{figure}[t!]
 \centering
 \includegraphics[width = \columnwidth]{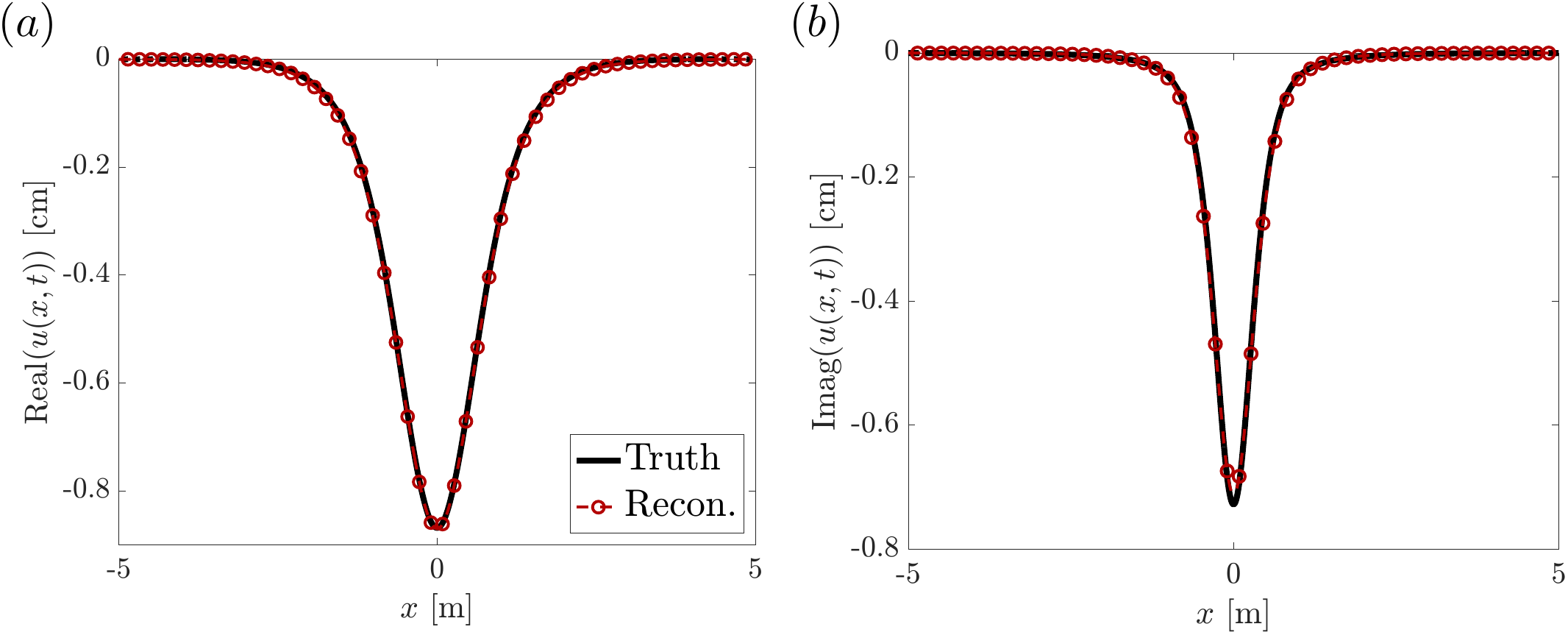}
 \caption{\label{fig:synth_env}Comparisons of the reconstructed envelope $\hat{u}$ from the LSTM prediction and the true envelope for the synthetic testing data. In (a) we show the real parts of $\hat{u}$ and $u$, and in (b) we present the imaginary parts. In both panels the solid black line corresponds to the ground truth, $u$, and the dashed red line with the circular markers corresponds to the reconstructed envelope $\hat{u}$.}
 \end{figure}
 
The procedure we outline above leads to excellent agreement between the LSTM predicted surface elevation $\hat{\eta}$ and the surface elevation obtained from this reconstruction $\hat{u}$, with relative errors less than 0.1\%.

In addition, if we compare the real and imaginary parts of $\hat{u}$ and $u$, which we present in Fig.~\ref{fig:synth_env}, we see an excellent agreement. There are some slight differences between the truth and the reconstruction which come from the error incurred by the LSTM's prediction of the full state. As mentioned previously, the relative error between the LSTM prediction, and the truth is about 2.0\%. The corresponding relative errors between the real and imaginary parts of $\hat{u}$ and $u$ are about 1.4\% and 3.4\% respectively, which indicates that the error from the reconstruction of the envelope is relatively small. 

The last piece of our framework is to forecast this reconstruction using NLS. We use the same parameters for pseudospectral truncation and the exponential time differencing scheme as above, and forecast to the final time $t = 30$. 
\begin{figure}
\centering
\includegraphics[width = \columnwidth]{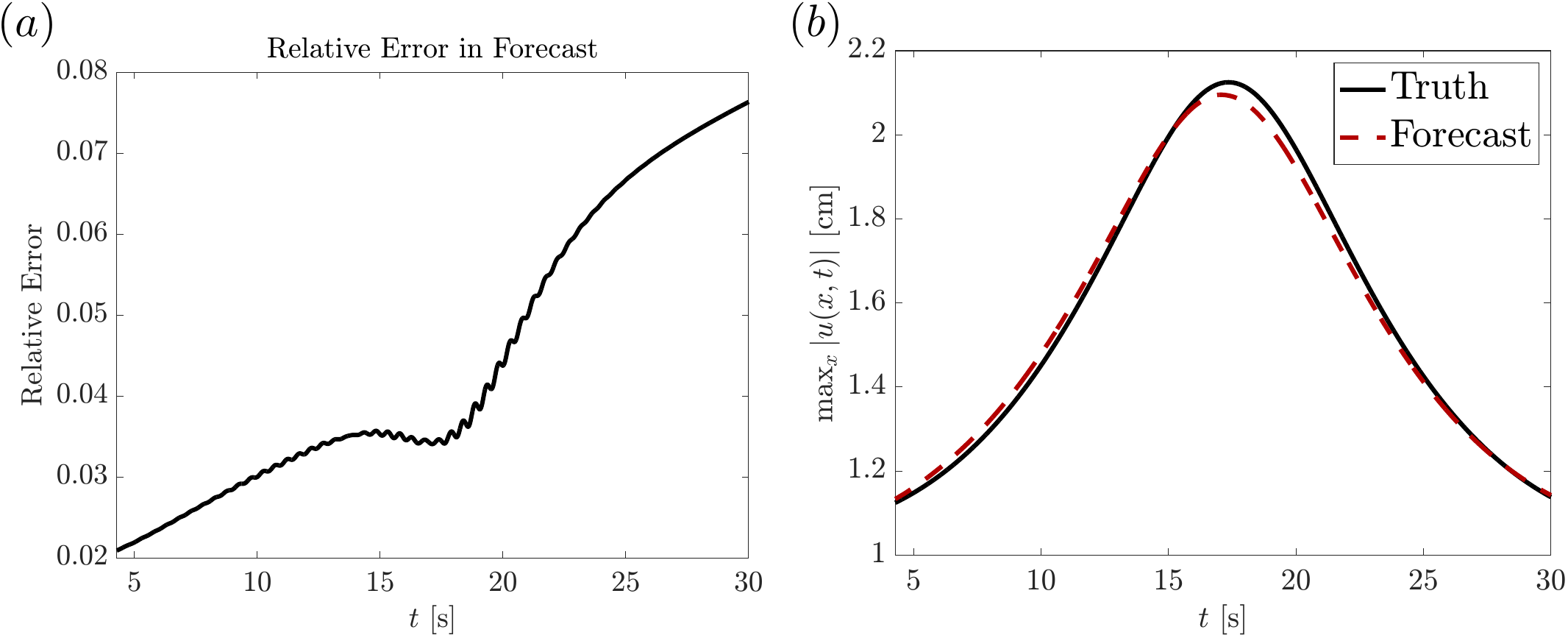}
\caption{\label{fig:synth_forecast}(a) The relative error in the forecast of the LSTM-reconstructed envelope. (b) The maximum value of the forecasted envelope (dashed red line) on top of the true maximum value of the true envelope (solid black line).}
\end{figure}
We present the results from the forecast in Fig.~\ref{fig:synth_forecast}. In terms of the relative error, we see that they are initially at 2\% and that they slowly grow to a maximum around 8\%. Qualitatively, however, we see that the results are very similar. More specifically, in Fig.~\ref{fig:synth_forecast} we can see that the timing of the formation of the peak of the rogue wave occurs with a very small delay, around 0.3 s, as compared to the truth. Moreover, the maximum amplitudes are very close with the forecast being about 0.03 cm smaller than the truth. The relative error in the maximum amplitude is approximately 1.8\%. Thus, for the synthetic data, the LSTM predictions are very accurate in the state estimate, the envelope reconstruction, and the forecast. 


\end{document}